\documentclass[preprint]{aastex}
\usepackage{graphicx}

\newcommand{\vB}{\mathbf{B}}

\newcommand{\ve}{\mathbf{e}}
\newcommand{\vv}{\mathbf{v}}

\newcommand{\grad}{\mbox{\boldmath$\nabla$}}

\newcommand{\diver}{\mbox{\boldmath$\nabla\cdot$}}
\newcommand{\vph}{\varphi}
\newcommand{\cm}{{\rm cm}}
\newcommand{\s}{{\rm s}}

\newcommand{\pd}{\partial}
\newcommand{\Pm}{P_{\rm m}}
\newcommand{\Rm}{R_{\rm m}}
\newcommand{\AR}{{\it A\kern-4pt\rm R}}
\begin{document}

\title{Magnetorotational instability of dissipative Couette flow}
\author{Jeremy Goodman and Hantao Ji}
\affil{Princeton University Observatory, Princeton, NJ 08544}
\email{jeremy@astro.princeton.edu}

\begin{abstract}
Global axisymmetric stability of viscous, resistive, magnetized
Couette flow is re-examined, with the emphasis on flows that would be
hydrodynamically stable according to Rayleigh's criterion:
opposing gradients of angular velocity and specific angular momentum.
In this regime, magnetorotational instabilities [MRI] may occur.
Previous work has focused on the Rayleigh-unstable regime.  To prepare
for an experimental study of MRI, which are of intense astrophysical
interest, we solve for global linear modes in a wide gap with
realistic dissipation coefficients.  Exchange of stability appears to
occur through marginal modes.  Velocity eigenfunctions of marginal
modes are nearly singular at conducting boundaries, but magnetic
eigenfunctions are smooth and obey a fourth-order differential
equation in the inviscid limit.  The viscous marginal system is of
tenth order; an eighth-order approximation previously used for
Rayleigh-unstable modes does not permit MRI.  Peak growth rates are
insensitive to boundary conditions.  They are predicted with
surprising accuracy by WKB methods even for the largest-scale mode.
We conclude that MRI is achievable under plausible experimental
conditions using easy-to-handle liquid metals such as gallium.

\end{abstract}

\section{Introduction}

To an even greater extent than large-scale terrestrial ones, astrophysical
flows are nearly inviscid.  Yet observations show that they dissipate
efficiently.  For example, accretion disks (flattened systems of
gas in orbit about a star or black hole) must lose orbital energy
in order that the gas flow onto the central object.  The influence of
turbulence has long been suspected, but purely hydrodynamic turbulence
is probably ineffective because of the strongly
stable radial angular momentum gradient in the disk.
When warm enough to be partially ionised, as they often are,  accretion
disks become magnetohydrodynamic (MHD) fluids.
It is now believed that turbulence and orbital decay are driven by
magnetorotational instabilities (MRI).
Although discovered by \cite{Velikhov} and
\cite{Chandra60}, MRI did not come to the attention of the
astrophysical community until rediscovered by \cite{BH91a}.  MRI
requires that the angular velocity ($\Omega$) must decrease with
distance from the axis, but it is distinguished from Rayleigh's
centrifugal instability by persisting when the gradient of specific
angular momentum is positive, \emph{i.e.} $\pd (r^4\Omega^2)/\pd r>0$.  A weak
background magnetic field is required, and MRI can be axisymmetric
when the field parallel to the rotation axis.  The rotation law of
accretion disks satisfies both conditions above (usually
$\Omega\propto r^{1/2}$), and while the magnetic geometry is
uncertain, it has little influence on the maximum MRI growth rate in
ideal MHD unless the field is purely toroidal \citep{BH92,TP96}.

There exists a body of experimental work on magnetized
Couette flow \citep{Donnelly60,Donnelly62,Donnelly64,Brahme70}, but the
MRI has never been demonstrated in the laboratory.  The main obstacle
is that liquid metals are strongly resistive on laboratory scales,
with magnetic diffusivity $\eta \gtrsim 10^3~\cm^2~\s^{-1}$.
The viscosity is much smaller, typically
$\nu\sim 10^{-3}~\cm^2~\s^{-1}$.  Their ratio 
is the magnetic Prandtl number $\Pm\equiv\nu/\eta\sim 10^{-6}$.

\cite{Chandra61} also analyzed \emph{dissipative} magnetized Couette
flow.  After laying out general equations for viscous and resistive
linearized axisymmetric perturbations, he discarded a term involving
shear from the azimuthal induction equation on the grounds that
$\Pm\ll 1$.  In fact the neglected term involves neither $\nu$ nor
$\eta$ directly.  Thus the commonly used name ``small-$\Pm$
approximation'' is somewhat unfortunate.  It presumes that viscous and
inertial forces are comparable throughout the flow.  Although this is
often the case for marginal Rayleigh instabilities resisted by viscous
(and perhaps magnetic) forces, it is not the case for the MRI modes of
interest here, where viscosity is important in boundary layers only.
We shall show that MRI modes do not exist in Chandrasekhar's
approximation (\S 3).\footnote{Of course he did find MRI modes, but in
a separate analysis assuming ideal MHD\citep{Chandra60}.}  We shall
introduce a different $\Pm\to0$ limit that retains all terms in the
induction equation but neglects viscous boundary layers.

Chandrasekhar's results have been refined by
\cite{ChangSart67,Hassard72,Vislovich86,TAS89,SAT94,ChenChang98}, but
always under Chandrasekhar's ``small-$\Pm$'' approximation.

In a previous paper \citep[henceforth Paper I]{JGK}, we have used
local WKB methods to survey the MRI regime for realistic materials
and laboratory parameters.  The most unstable modes (and perhaps the
only ones accessible to experiment in the near term) have wavelengths
twice as large as the apparatus, so that WKB methods are not to be
trusted \emph{a priori}.  The present paper therefore discusses the
global linear analysis.  We integrate the full set of viscous and
resistive equations \emph{via} an initial-value scheme to obtain
numerical growth rates for cases that would be stable by Rayleigh's
criterion.  Boundary conditions are problematic for inviscid marginal
modes.  Nevertheless,  the locally obtained growth
rates are found to be good approximations, even though the radial
eigenfunctions are far from being sinusoidal.
We predict instability under feasible conditions:
gap widths and heights of order ten centimeters, fields strengths of
several kilogauss, and rotation rates of several hundred radians per
second.

A supplementary analysis shows
that curves of marginal MRI stability are well approximated by
completely neglecting viscous terms, especially for insulating
cylinders.  Although $\Pm\to0$, this is \emph{not} Chandrasekhar's
approximation.  It eliminates the velocity perturbations from the
analysis and leads to a system of only four radial derivatives.  The
boundary conditions on velocity are not satisfied by the reduced
system even with stress-free (slipping) boundaries.  This is resolved
by restoring the viscosity but there is little change in the magnetic
eigenfunctions or the growth rate when $\Pm$ is realistically small.
Hence when MRI are of primary interest,
dimensionless numbers based on viscosity,
\emph{viz.} Hartmann and Taylor numbers, are less useful for
characterizing stability than numbers that remain finite as $\Pm\to0$,
such as Lundquist number and magnetic Reynolds number.

\section{Basic equations and boundary conditions}

We use cylindrical coordinates $r\theta z$ aligned with the rotation of
the fluid, and gaussian electromagnetic units.
In equilibrium, the magnetic field is constant and parallel
to the axis; it is described by the associated Alfv\'en speed
$V_A\equiv B_0/\sqrt{4\pi\rho}$, where $\rho$ is the (constant) density
of the liquid metal.  The angular velocity is $\Omega(r)$.
Perturbations are axisymmetric
and sinusoidal in $z$ with wavelength $2\pi/k$:
\begin{equation}\label{zdepend}
\begin{array}{lll@{\hspace{20pt}}lll}
\delta B_r/\sqrt{4\pi\rho} &=& \beta_r(r,t)\cos kz 
&\delta v_r &=& \vph_r(r,t)\sin kz \\
\delta B_\theta/\sqrt{4\pi\rho} &=& \beta_\theta(r,t)\cos kz 
&\delta v_\theta &=& \vph_\theta(r,t)\sin kz \\
\delta B_z/\sqrt{4\pi\rho} &=& \beta_z(r,t)\sin kz 
&\delta v_z &=& \vph_z(r,t)\cos kz
\end{array}
\end{equation}
We often write $h\equiv\pi/k$ for half the wavelength, having in mind
an experiment of finite height $h$ and rigid vertical boundaries.

Perturbations associated with a single mode
are exponential in time, but it is convenient to allow for
general time dependence and discover the fastest-growing mode by integrating
the linearized equations forward in time.  These equations are (see the
Appendix)
\begin{eqnarray}
\label{vsystem1}
\dot\beta_\theta &=& \eta(D-k^2)\beta_\theta 
+kV_A\vph_\theta +\underline{r\Omega'\beta_r}\\[5pt]
\label{vsystem2}
\dot\vph_\theta&=& \nu(D-k^2)\vph_\theta -kV_A\beta_\theta
\,-r^{-1}(r^2\Omega)'\vph_r\\
\label{vsystem3}
\dot\beta_r &=& \eta(D-k^2)\beta_r+kV_A\vph_r\\
\label{vsystem4}
\dot\vph_r&=&\nu(D-k^2)\vph_r
-kV_A\beta_r \,+\Pi\\
\label{vsystem5}
(k^2-D)\Pi &=& 2\Omega k^2\vph_\theta
\end{eqnarray}
The prime in eqs.~(\ref{vsystem1})-(\ref{vsystem2}) means $\pd/\pd r$, and
\begin{displaymath}
D\equiv \frac{\pd^2}{\pd r^2} +\frac{1}{r}\frac{\pd}{\pd r} - \frac{1}{r^2}.
\end{displaymath}
Note that \cite{Chandra61} denotes the latter operator by $DD_*$.

Equations (\ref{vsystem1})-(\ref{vsystem2}) 
are the azimuthal components of the
induction and euler equations, while (\ref{vsystem3})-(\ref{vsystem4}) 
are the corresponding radial components.
By eliminating the auxiliary function $\Pi$
between eqs.~(\ref{vsystem4})-(\ref{vsystem5}), one obtains an equation
equivalent to Chandrasekhar's eq.~(168), although
some differences in sign occur because we
have taken perturbations proportional to $\sin kz$
where he took $\cos kz$ and {\it vice versa}.
Our eqs.~(\ref{vsystem1}), (\ref{vsystem2}), and (\ref{vsystem3})
correspond to his eqs.~(163), (160), and (162), respectively.

The flow is confined between concentric cylinders with radii $r_1<r_2$.
If these are perfectly conducting, the magnetic boundary conditions are
\begin{equation}\label{conducting}
\beta_r = 0,\hspace{60pt} (r\beta_\theta)' = 0.
\end{equation}
If perfectly insulating, then ($I_n$ and $K_n$ are modified 
Bessel functions)
\begin{eqnarray}\label{insulating}
\frac{\pd}{\pd r}(r\beta_r) &=& \beta_r\times
\cases{ [krI_0(kr)]/I_1(kr)\vphantom{A_0'\over B_0'} & at $r=r_1$\cr
-[krK_0(kr)]/K_1(kr)\vphantom{A_0'\over B_0'} & at $r=r_2$\cr}\nonumber
\\[10pt]
\beta_\theta &=& 0.
\end{eqnarray}
The conditions on velocity are
\begin{equation}
\vph_r=0,\hspace{40pt}\nu\vph_\theta=0=\nu(r\vph_r)'.
\end{equation}
We have put the viscosity in the latter two conditions so that
$\vph_\theta$ and $\vph_z=(r\vph_r)'/(kr)$ will be unconstrained
when the flow is inviscid.

The insulating conditions (\ref{insulating}) are not accurate for an
experiment of finite height $h=\pi/k$, since they assume a vertically
periodic solution for the vacuum field outside the cylinders.  The
conducting conditions (\ref{conducting}) do not have this drawback.
In both cases, there will be viscous boundary layers at the top and
bottom (unless the end caps rotate differentially), but we expect that
the error committed by neglecting those layers is small for $\Pm\sim
10^{-6}$ and growth times shorter than the Ekman circulation time.
In any event, the end caps should be insulating so that no
magnetic stress acts upon them.

\section{Why the small-$\Pm$ approximation suppresses the MRI}

The underlined term in eq.~(\ref{vsystem1}) is the critical
one that \cite{Chandra61} and subsequent authors neglected on the
grounds that $\Pm\ll 1$.
To see that this term is necessary to the MRI, it is useful to reduce
eqs.~(\ref{vsystem1})-(\ref{vsystem5}) to a single equation.

For brevity's sake,  assume a mode with definite
growth rate $s$, and write $D_k\equiv D-k^2$, $\omega_A\equiv kV_A$.
Eq.~(\ref{vsystem3}) yields
\begin{equation}\label{phirsol}
\vph_r= \omega_A^{-1}(s-\eta D_k)\beta_r\,.
\end{equation}
Substituting for $\vph_r$ in eq.~(\ref{vsystem4}), applying
$D_k$, and eliminating $D_k\Pi$ \emph{via} eq.~(\ref{vsystem5}), one has
\begin{equation}\label{phitsol}
\vph_\theta=-(2\Omega k^2\omega_A)^{-1}\left[(s-\nu D_k)(s-\eta D_k)
+\omega_A^2\right]D_k\beta_r\,.
\end{equation}
Solving for $\beta_\theta$ from eq.~(\ref{vsystem2}) and eliminating
$\vph_\theta$ and $\vph_r$ in favor of $\beta_r$, one has
\begin{eqnarray}\label{btsol}
\beta_\theta&=&\omega_A^{-2}\left\{(s-\nu D_k)\frac{1}{2\Omega k^2}\left[
(s-\nu D_k)(s-\eta D_k)+\omega_A^2\right]D_k\right.\nonumber\\
&~&\left. -\frac{(r^2\Omega)'}{r} (s-\eta D_k)\right\}\beta_r\,.
\end{eqnarray}
Using these to eliminate $\vph_\theta$ and $\beta_\theta$
from eq.~(\ref{vsystem1}) yields the desired tenth-order equation:
\begin{eqnarray}\label{tenthorder}
&&\left\{\left[(s-\nu D_k)(s-\eta D_k)+\omega_A^2\right]\frac{1}{2\Omega}
\left[(s-\nu D_k)(s-\eta D_k)+\omega_A^2\right](-k^{-2}D_k)\right.\nonumber\\
&& \qquad\qquad
\left. +(s-\eta D_k)\frac{(r^2\Omega)'}{r} (s-\eta D_k)\right\}\beta_r
~=~ -\omega_A^2\,\underline{r\Omega'\,\beta_r}.
\end{eqnarray}

We are interested in Rayleigh-stable cases.  Without loss of
generality, we may assume that the angular velocity ($\Omega$) and
vorticity [$r^{-1}(r^2\Omega)'$] are positive throughout the flow, but
the shear ($r\Omega'$) may be negative.  Now $D_k$ is clearly negative
definite and self adjoint with either of the boundary conditions
(\ref{conducting}) or (\ref{insulating}).  In the narrow-gap limit
$(r_2-r_1)/(r_2+r_1)\to0$, the angular velocity and vorticity are
positive constants.  It follows that the operator in braces in the
lefthand side of eq.~(\ref{tenthorder}) is positive definite for
nonnegative real $s$.  Therefore, {\it at least in the narrow-gap limit,
there can be no modes with positive real growth rate when the
underlined shear term is neglected and the Rayleigh stability
criterion is satisfied}.  We interpret this to mean that the MRI is
not present.

Previous studies of the time-dependent
problem have neglected the time derivatives in the induction equation
as well as the underlined term \citep{ChangSart67,ChenChang98}; in
this case, the operator in question becomes quadratic in $s$ so that
the coefficient of $\mbox{Imag}(s)$ is positive-definite for
$\mbox{Real}(s)>0$, which rules out complex growing modes, \emph{i.e.}
overstabilities.

Unfortunately, we cannot draw such strong conclusions in the wide-gap
case where $\Omega$ and perhaps also $r^{-1}(r^2\Omega)'$ vary
significantly with radius.  The operator on the left side of
eq.~(\ref{tenthorder}) is no longer self adjoint in general,
because $D_k$ and $\Omega$ do not commute.
For marginal modes, however, 
\begin{equation}\label{eigthorder}
\left\{\left[\nu D_k^2+\frac{\omega_A^2}{\eta}\right]\frac{\eta}{2\Omega k^2}
\left[\nu D_k^2+\frac{\omega_A^2}{\eta}\right]~-~
\eta D_k\frac{(r^2\Omega)'}{r}\right\}\vph_r
~=~ - \omega_A\,\underline{r\Omega'\,\beta_r}\,,
\end{equation}
in which we have used eq.~(\ref{phirsol}).
In this case the problem is only eighth order in radial derivatives
if the righthand side is neglected, as noted by \cite{Chandra61}.
More importantly, the operator in these curly braces \emph{is} self adjoint
in the interesting special case that
\begin{equation}\label{Omegaprof}
\Omega(r)= a + \frac{b}{r^2},
\end{equation}
since $r^{-1}(r^2\Omega)'=2a$ is then constant, and the rest of
the operator is symmetrical.
Eq.~(\ref{Omegaprof}) is the angular-velocity profile of a Couette
flow in steady state, because it implies a radially constant viscous 
angular-momentum flux.  We conclude that there are no
Rayleigh-stable \emph{marginal} modes when the magnetic shear term is
neglected, even for wide gaps.

\section{Marginal inviscid modes}\label{sec_inviscid}

If $\nu=0$, then 
as the growth rate $s\to0$, eq.~(\ref{tenthorder}) reduces to
\begin{equation}\label{fourthorder}
\left[\eta^2 D_k\,\frac{(r^2\Omega)'}{r}
~-\,\frac{\omega_A^4}{2\Omega k^2}\right]D_k\beta_r~=~-\omega_A^2
\underline{r\Omega'\,\beta_r}\,.
\end{equation}
Even with the righthand side included, this is only a fourth-order system.
Paradoxically, there are six boundary conditions to be satisfied:
$\vph_\theta$ and $\vph_z$ are not constrained when $\nu=0$, but
$\vph_r=0$ at both boundaries, in addition to a total of four magnetic
conditions.  For the
insulating case (\ref{insulating}), the paradox is resolved because
eqs.~(\ref{phirsol}) and (\ref{btsol}) show that both $\vph_r=0$ and
$\beta_\theta=0$ are equivalent to $D_k\beta_r=0$, so that
there are only four independent boundary conditions after all.
But in the conducting case (\ref{conducting}), we have \emph{via}
eqs.~(\ref{phirsol})-(\ref{btsol}) that 
$\beta_r=D_k\beta_r=(rD_k\beta_r)'=0$ at both boundaries,
and not all six conditions can be satisfied.

The crux of the difficulty is the azimuthal euler equation
(\ref{vsystem2}), which reduces to an algebraic relation. 
At zero viscosity and growth rate, azimuthal force balance requires
\begin{equation}\label{forcebalance}
2\Omega\delta v_r = \frac{B_0}{4\pi\rho}\frac{\pd\delta B_\theta}
{\pd z}~,
\end{equation}
so that the Lorentz force ($\delta j_r \times B_0$) balances the Coriolis
force, which vanishes at the boundary.  If
the boundary is insulating, then
$\delta B_\theta$ ($\propto\delta j_r$) also vanishes.
But at conducting boundaries, $\delta B_\theta\ne0$
($\delta j_r \ne 0$), so that viscosity
must intervene to maintain azimuthal
force balance. For small $\Pm$, the marginal eigenfunction displays
a dramatic boundary layer (Fig.~\ref{fig_cmode}).

Viscous boundary layers are common in hydrodynamics.  Normally they
occur because the tangential component of velocity must match that of
the boundary itself, even when the viscosity is small (a ``no-slip''
boundary condition).  For conducting cylinders, however, {\it a
boundary layer would occur even if the viscous stress vanished at the
boundary}, because of the impossibility of satisfying
eq.~(\ref{forcebalance}).  In the present case, 
the viscous layer is driven by tangential \emph{magnetic field}
(or normal component of current) rather than tangential velocity.

To summarize, the inviscid limit is singular for marginally stable
modes.  The eigenfunctions become ill-behaved because there are more
boundary conditions than radial derivatives to satisfy them, at least
for conducting boundaries.  The numerical evidence presented below
indicates, however, that the locus of marginal stability in the
parameter space of equilibria is actually continuous as $\Pm\to0$.
Hence the relatively simple fourth-order differential equation
(\ref{fourthorder}) predicts marginal stability reasonably well when
$\Pm$ is sufficiently small.

\section{Numerical results}

We have approximated eqs~(\ref{vsystem1})-(\ref{vsystem5}) by
finite-difference 
equations on a radial grid uniformly spaced in $x\equiv\ln r$.
The background angular velocity has the form (\ref{Omegaprof}),
since this is easiest to realize experimentally.
The grid spacing $\Delta x$ must be chosen fine enough so that
$\Delta r^2/\nu$ is smaller than the interesting physical timescales
in the problem, \emph{viz.} $\omega_A^{-1}$ and $(\eta d^2)^{-1}$
(where $d\equiv r_2-r_1$ is the gap width), otherwise the viscous
boundary layer will not be resolved.
The minimum number of grid cells $N\sim\Pm^{-1/2}\sim 10^3$.
Our scheme has second-order spatial accuracy, even at the
the boundaries.

To ensure numerical stability, we use fully implicit time differencing.
Spatial differences are written in terms of
the unknown dependent variables at the new time step, so that a
large linear system must be solved.
Actually, our finite-difference matrix is band diagonal with 10 nonzero
codiagonals in each of $5N$ rows, and it is independent of time step.
We perform $LU$ decomposition
at the start of the evolution so that only the
back substitution must be performed anew at each step.

When a growing mode is present, it eventually dominates.  Then
the perturbation in radial magnetic field
at successive time steps $t_n$ and $t_{n+1}=t_n+\Delta t$
are related by, for example,
\begin{equation}\label{timedifference}
(1-\hat s\Delta t)\,\hat\beta_r(x_j,t_{j+1})=\hat\beta_r(x_j,t_{j}),
\end{equation}
if $\hat s>0$ is the appropriate eigenvalue of the matrix defined
by the spatial difference scheme and therefore an estimate of the physical
growth rate, and $\hat\beta_r$ is the corresponding eigenvector.
Given $\hat\beta_r(x_j,t_j)$ and $\hat\beta_r(x_j,t_{j+1})$, we can
compute $\hat s$ from eq.~(\ref{timedifference})
without any truncation error in $\Delta t$ as long as $\hat s\Delta t<1$.
The eigenfunctions are similarly independent of $\Delta t$.
This is advantageous
close to marginal stability where $\hat s$ is small.
Our procedure is equivalent to  
finding the most-positive eigenvalue
of the time-evolution matrix by inverse iteration.  By extending 
eq.~(\ref{timedifference}) to a three-term recurrence relation, we have
allowed for complex eigenvalues.
But in practice, all of our growing modes appear to have purely
real values of $\hat s$.
Of course, our method is not immune to \emph{spatial} truncation
errors; these are $O(\Delta x^2)$ because we use second-order
spatial differencing.

\begin{figure}
\scalebox{0.6}{\includegraphics{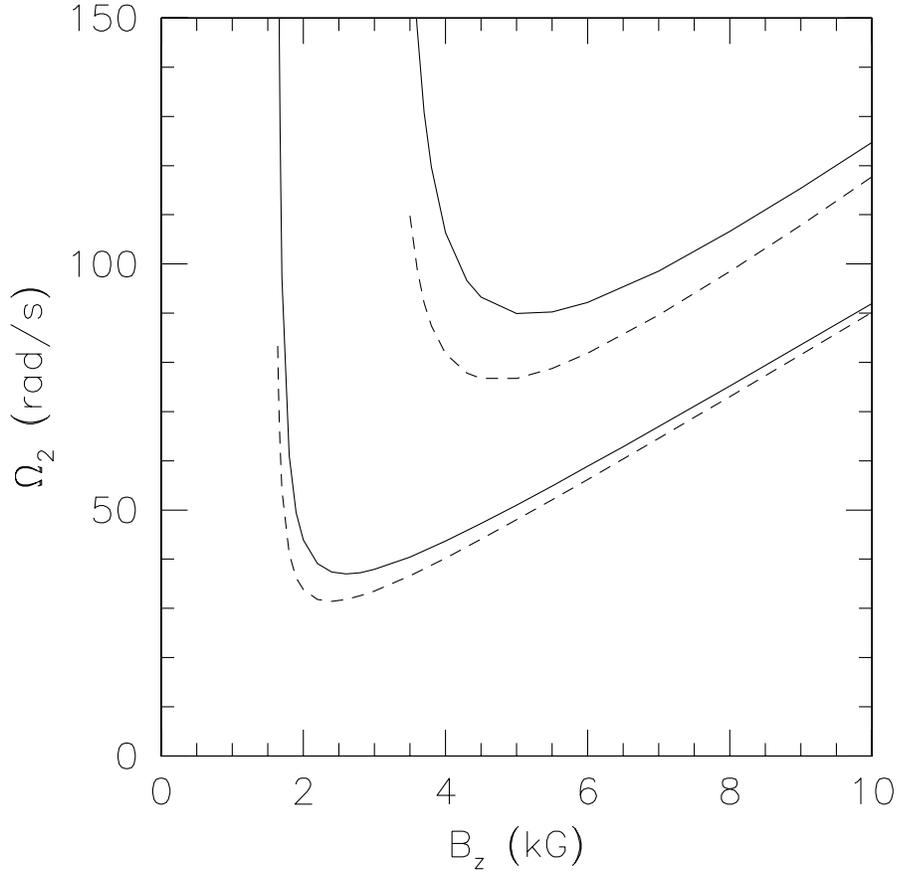}}
\caption{Marginal stability for liquid gallium Couette flow between
conducting cylinders of radii
$r_1=5\,\cm$, $r_2=15\,\cm$, and height $h=10\,\cm$.
Solid lines computed from eqs.~(\ref{vsystem1})-(\ref{vsystem5});
dashed from inviscid approximation 
(\ref{inviscid1})-(\ref{inviscid2}).
Lower curves for dimensionless vorticity $\zeta=2/11$, 
upper ones for $\zeta=4/7$. Instability occurs above the curves.
In dimensionless parameters,
$S\approx 0.92(B/10\,kG)$; and 
$\bar\Rm\approx 0.66(\Omega_2/100\,\mbox{rad s}^{-1})$ (upper curves),
$\bar\Rm\approx 0.73(\Omega_2/100\,\mbox{rad s}^{-1})$ (lower).
\label{fig_mc}}
\end{figure}

\begin{figure}
\scalebox{0.6}{\includegraphics{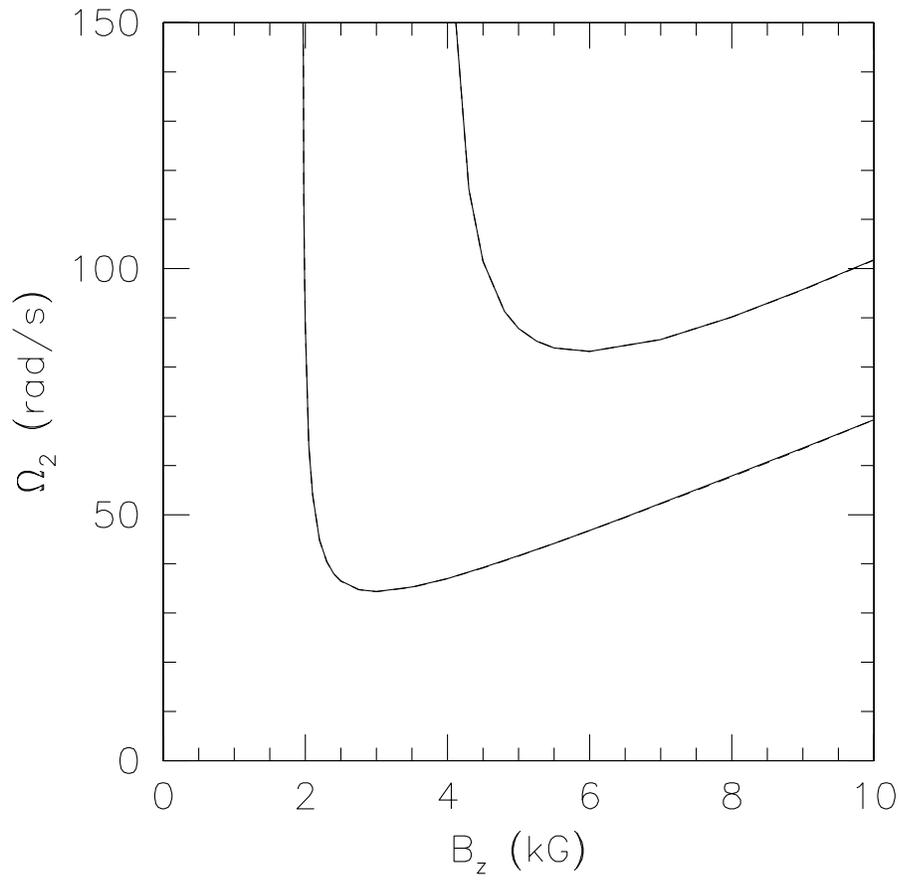}}
\caption{Like Fig.~(\ref{fig_mc}), but for insulating cylinders
(\ref{insulating}).
Viscous and inviscid results differ by less than the line thickness.
\label{fig_mi}}
\end{figure}

Our initial-value code can treat slightly stable cases as well as
unstable ones.  By interpolation, we find parameters for
marginal stability.  Results are shown in Figs.~(\ref{fig_mc}) \&
(\ref{fig_mi}) for material properties approximating liquid gallium,
\emph{viz.} $\rho=6\,\mbox{g cm}^{-3}$, 
$\eta=2000\,\cm^2\s^{-1}$ and $\Pm=1.6\times 10^{-6}$.

Marginal stability defines one constraint among the eight parameters
defining the Couette flow: $\eta$, $\nu$, $r_1$, $r_2$, $k$,
$\Omega_1$, $\Omega_2$, and $V_A$.
Six independent dimensionless combinations of these
can be formed. The magnetic Prandtl number $\Pm\equiv\nu/\eta$ is one of
these.  The aspect ratio
\begin{equation}\label{aspect}
A \equiv \frac{r_2+r_1}{d}~,
\end{equation}
and the elongation of the toroidal cross section,
\begin{equation}
\epsilon \equiv \frac{h}{d}
\end{equation}
define the geometry, where $d\equiv r_2-r_1$ is the gap width and
$h\equiv \pi/k$ is half the vertical wavelength.
It turns out that the magnetic eigenfunctions $\beta_r$ and $\beta_\theta$
of the most unstable mode have a roughly parabolic dependence on $r$,
and since one or the other vanishes at both boundaries, the effective
horizontal wavenumber is $\approx \pi/d$.  The total wavenumber is then
\begin{equation}\label{Kdef}
K \equiv\left(k^2 +\frac{\pi^2}{d^2}\right)^{1/2}=\frac{\pi}{h}
\sqrt{1+\epsilon^2}.
\end{equation}
The Lundquist number
\begin{equation}\label{Sdef}
S\equiv \frac{kV_A}{\eta K^2} = \frac{V_A h}{2\pi\eta}\,\frac{2}{\epsilon^2+1}
\end{equation}
scales the Alfv\'en frequency against the magnetic diffusion rate $\eta K^2$.
In astrophysics, $S$ is often called ``magnetic
Reynolds number.''  The plasma community generally reserves the
latter term for a quantity involving fluid velocity,
so we define the local magnetic Reynolds number by
\begin{equation}\label{Rmdef}
\Rm(r) \equiv \frac{\Omega}{\eta K^2}.
\end{equation}
The viscous Reynolds number is of course $\Rm/\Pm$.
The dimensionless vorticity
\begin{equation}\label{zetadef}
\zeta(r)\equiv \frac{(r^2\Omega)'}{r\Omega}
\end{equation}
parametrizes the angular momentum gradient,
and the Rayleigh stability criterion is simply $\zeta(r)\ge 0$.
In the astrophysical
literature, the radial variation of angular velocity is often described
by an index
\begin{displaymath}
q\equiv -\frac{d\ln\Omega}{d\ln r}\qquad\mbox{so that}\qquad \zeta=2-q.
\end{displaymath}
Of course, $\zeta=2$ and $q=0$ in a uniformly rotating flow.

When the aspect ratio is modest, $\Rm$ and $\zeta$ 
may vary considerably across the gap, and it is useful to define
mean values of these dimensionless parameters.
Following Paper I, 
we introduce $\bar\Omega\equiv\sqrt{\Omega_1\Omega_2}$ and
\begin{equation}\label{meanvals}
\bar\Rm\equiv\frac{\bar\Omega}{\eta K^2},\qquad
\bar\zeta\equiv\frac{2(r_2^2\Omega_2-r_1^2\Omega_1)}
{(r_2^2-r_1^2)\bar\Omega}
\end{equation}

The locus of marginal stability is actually a hypersurface in
the space $(\Pm, A ,\epsilon,\bar\zeta,S,\bar\Rm)$.  The
curves in Figs.~(\ref{fig_mc})-(\ref{fig_mi}) are cuts through this
locus at constant values of the first four parameters:
$\Pm=0$ and $\Pm=1.6\times 10^{-6}$;
$ A =\epsilon=1$; and two positive values of $\zeta$ as indicated.
The curves are drawn in physical units for the density and
diffusivity of gallium.

The inviscid results shown in
in these figures were calculated by an independent
numerical method based on eq.~(\ref{fourthorder}), which 
we have unpacked into a pair of second-order equations 
[using eq.~(\ref{btsol}) with $s,\nu\to0$]:
\begin{eqnarray}
D_k\beta_r &=& \frac{r\omega_A^2}{\eta (r^2\Omega)'}\,\beta_\theta
\qquad\qquad\qquad\quad~=
~K^2\,\frac{S^2}{\zeta\Rm}\,\beta_\theta,\label{inviscid1}\\[10pt]
D_k\beta_\theta &=& \frac{\omega_A^4\,r}{2\Omega(r^2\Omega)'\,\eta^2k^2}
\,\beta_\theta ~-\, \frac{r\Omega'}{\eta}\beta_r
~=~K^2\left[\frac{(1+\epsilon^2)S^4}{2\zeta\Rm^2}\,\beta_\theta
+(2-\zeta)\Rm\beta_r\right],\label{inviscid2}
\end{eqnarray}
together with the magnetic boundary conditions (\ref{conducting}) or
(\ref{insulating}).
Because of the large magnetic diffusivity, the magnetic
variables are very well-behaved, so that this fourth-order system
can be solved efficiently by a shooting method.

\begin{figure}
\hspace*{-0.5in}
\begin{minipage}{3in}
\scalebox{0.45}{\includegraphics{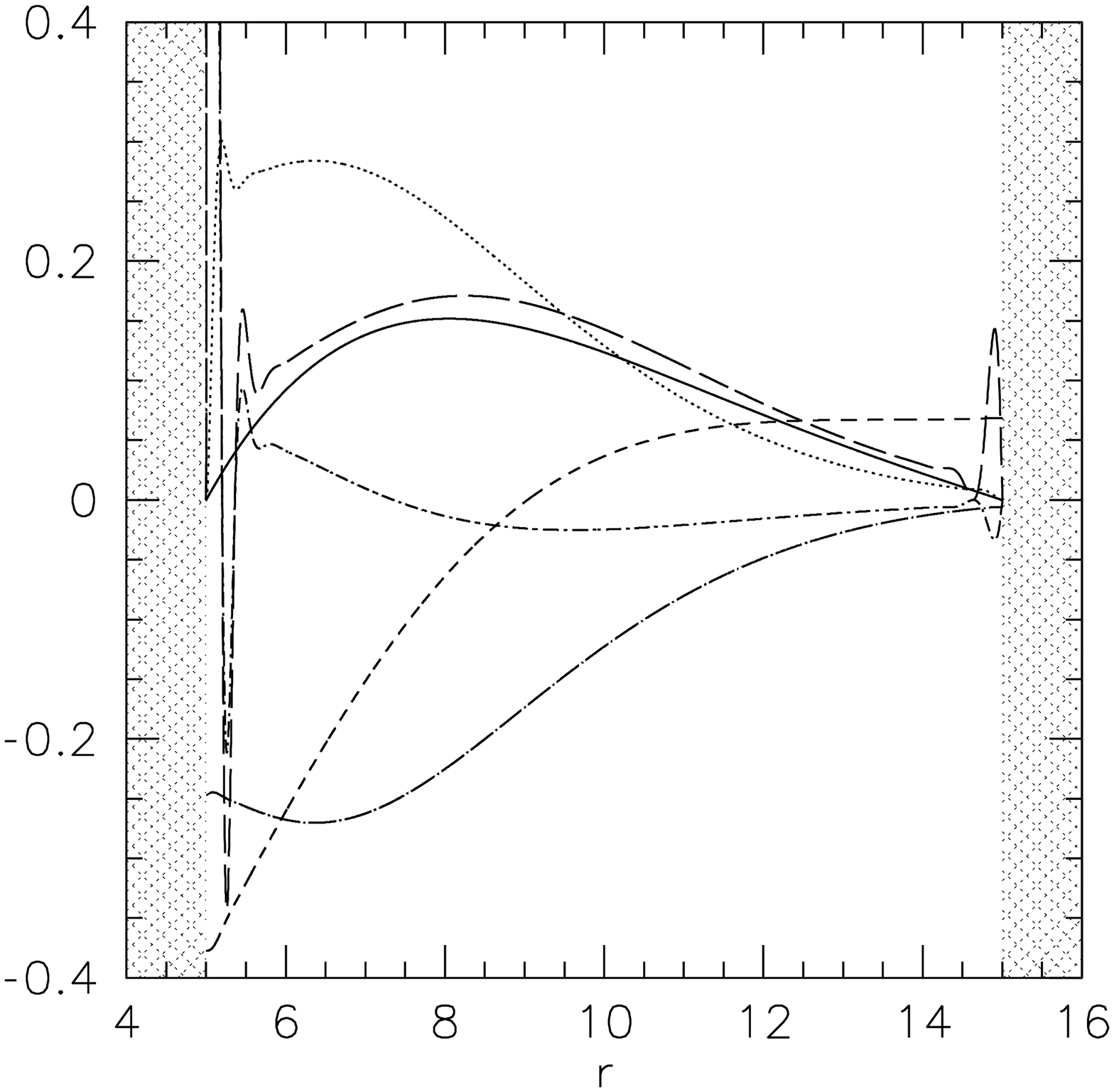}}
\end{minipage}
\hfill
\begin{minipage}{3in}
\scalebox{0.45}{\includegraphics{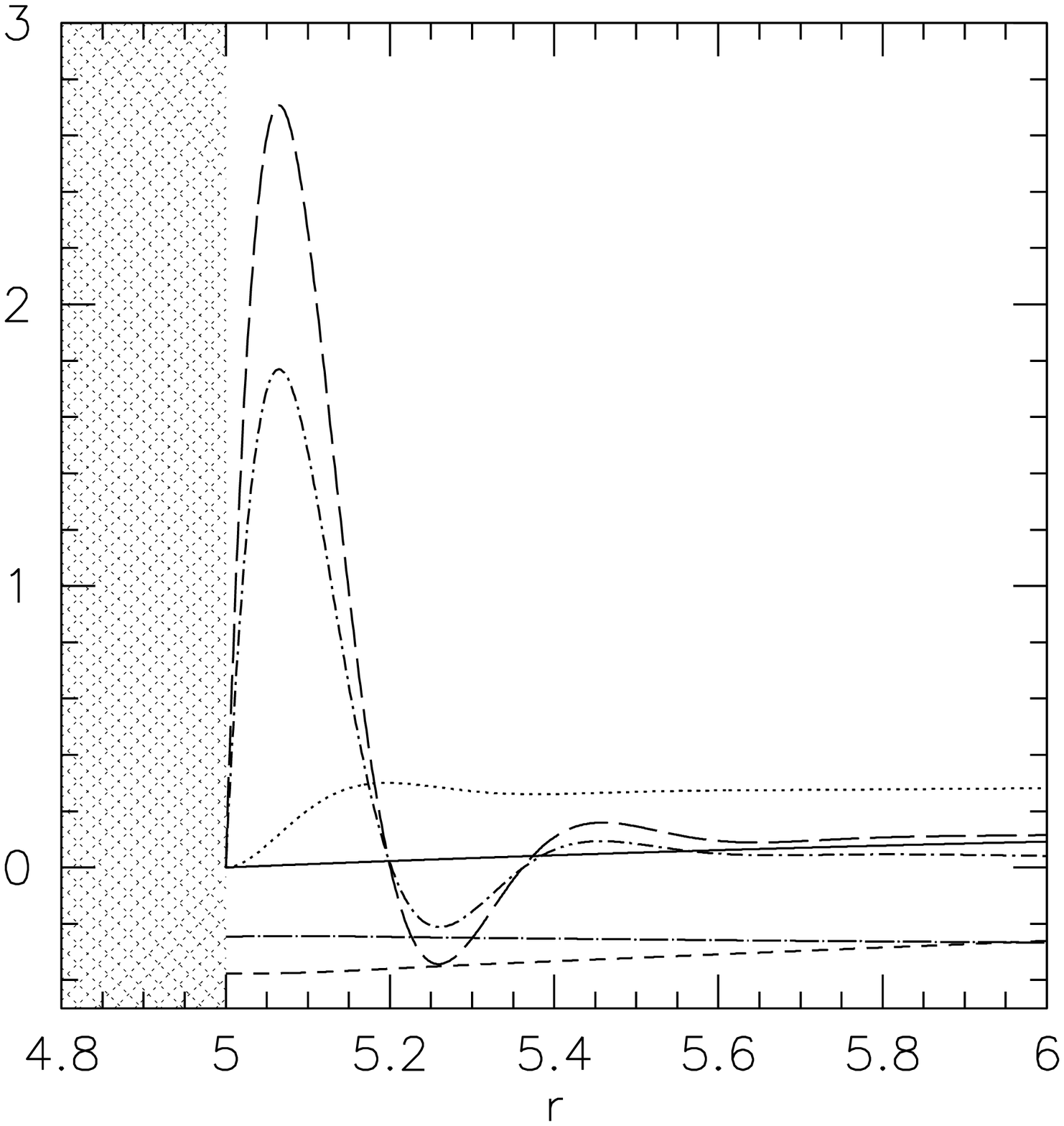}}
\end{minipage}
\caption{Global and closeup views
of marginal eigenmode with conducting boundaries. $B=3\,\mbox{kG}$, 
$\Omega_1=314.$, $\Omega_2=37.9~{\rm rad~s^{-1}}$
\& $\bar\zeta=0.0632$.
Solid curve: $\beta_r$.
Short-dashed: $\beta_z$
Dot-long-dashed: $\beta_\theta\times 5$.
Dotted: $\vph_r\times 1/3$.
Long-dashed: $\vph_\theta$.
Dot-short-dashed: $\vph_z\times0.07$.
\label{fig_cmode}}
\end{figure}

\begin{figure}
\hspace*{-0.5in}
\begin{minipage}{3in}
\scalebox{0.45}{\includegraphics{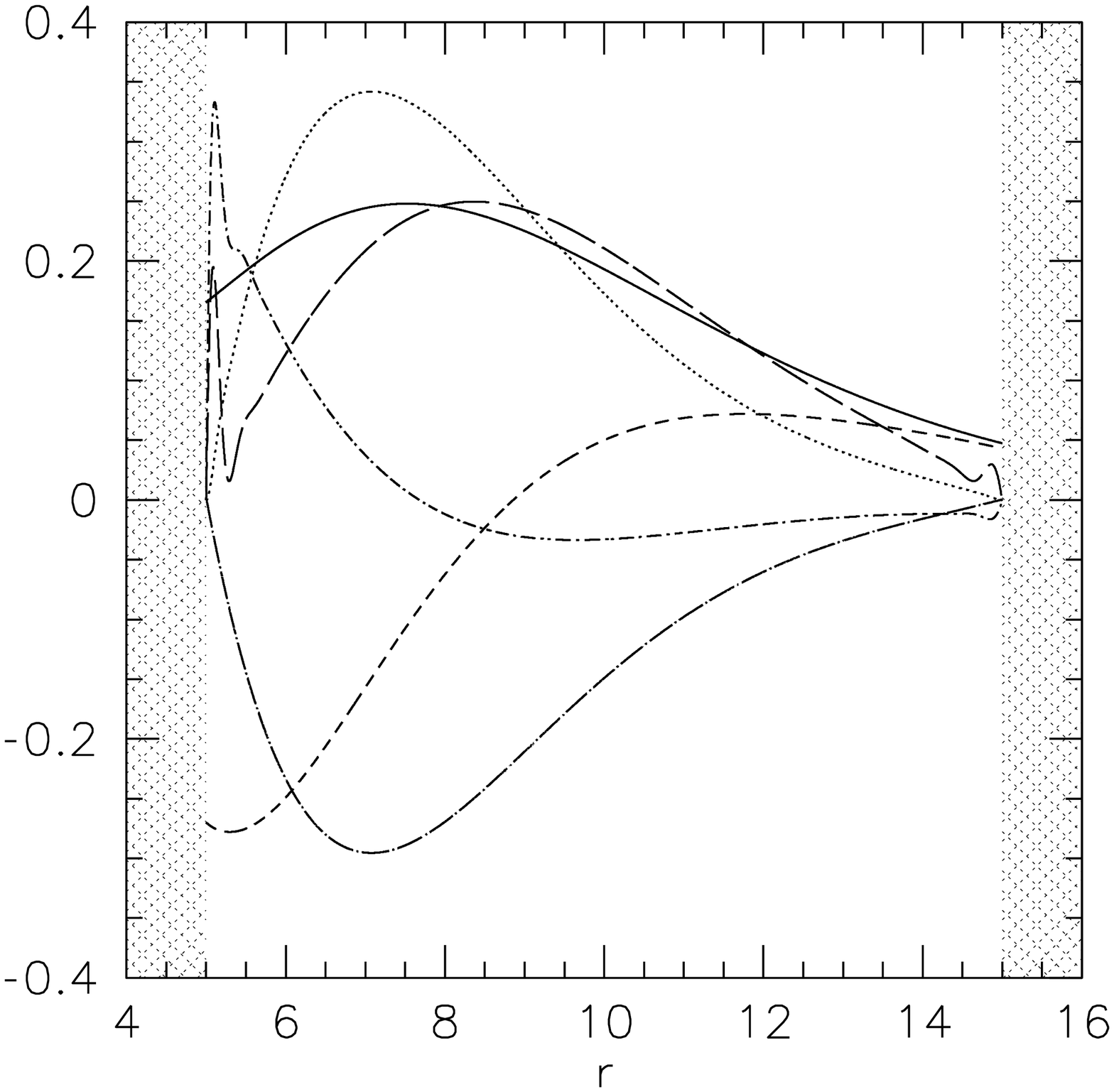}}
\end{minipage}
\hfill
\begin{minipage}{3in}
\scalebox{0.45}{\includegraphics{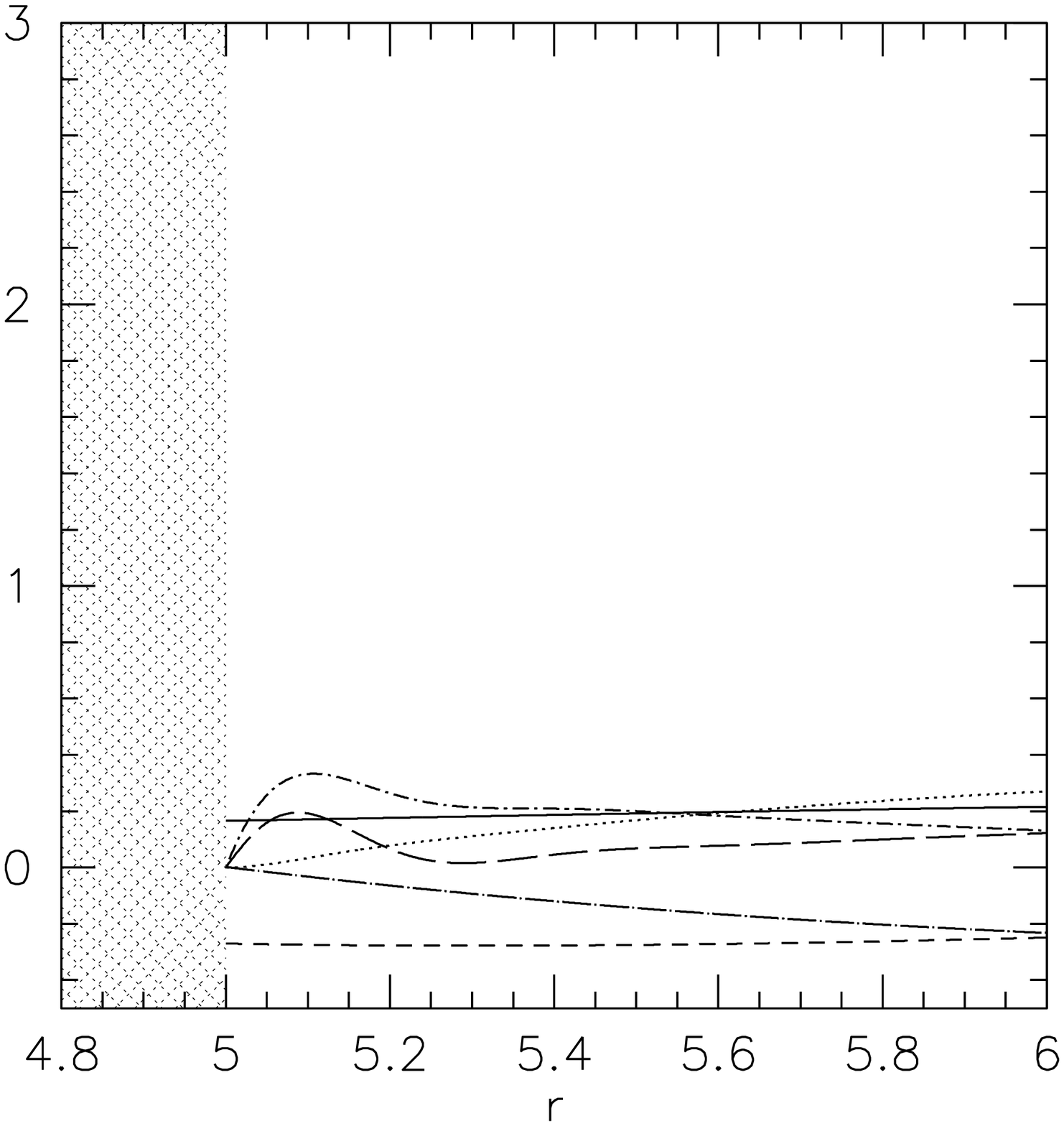}}
\end{minipage}
\caption{Like Fig.~\ref{fig_cmode} but for insulating boundaries
and $\Omega_1=284.$, $\Omega_2=34.4~{\rm rad~s^{-1}}$, $\bar\zeta=0.0632$.
\label{fig_imode}}
\end{figure}

Comparing Figs.~\ref{fig_cmode} and \ref{fig_imode}, one sees that
viscosity is more important for conducting boundary conditions than for
insulating ones.  In the conducting case, $\vph_r$ and
$\beta_\theta$ are nearly proportional to one another throughout most
of the flow [Fig.~\ref{fig_cmode}].
 This follows from eq.~(\ref{vsystem2}) in the limit
$s,\nu\to 0$, as we discussed in \S\ref{sec_inviscid}.  Since the
radial velocity perturbation ($\propto\vph_r$) must vanish at the wall
but the azimuthal magnetic perturbation ($\propto\beta_\theta$) does
not, there is a thin boundary layer in which viscous stress balances
the azimuthal magnetic force.  The righthand panel shows that the
boundary layer is well resolved by these calculations, which use 4000
grid points uniformly spaced in $\ln r$ across the gap.  At an
insulating boundary, on the other hand, $\beta_\theta$ vanishes with
$\vph_r$, and this leads to a much less dramatic viscous layer
(Fig.~\ref{fig_imode}).

\begin{figure}
\hspace*{-0.5in}
\begin{minipage}{3in}
\scalebox{0.45}{\includegraphics{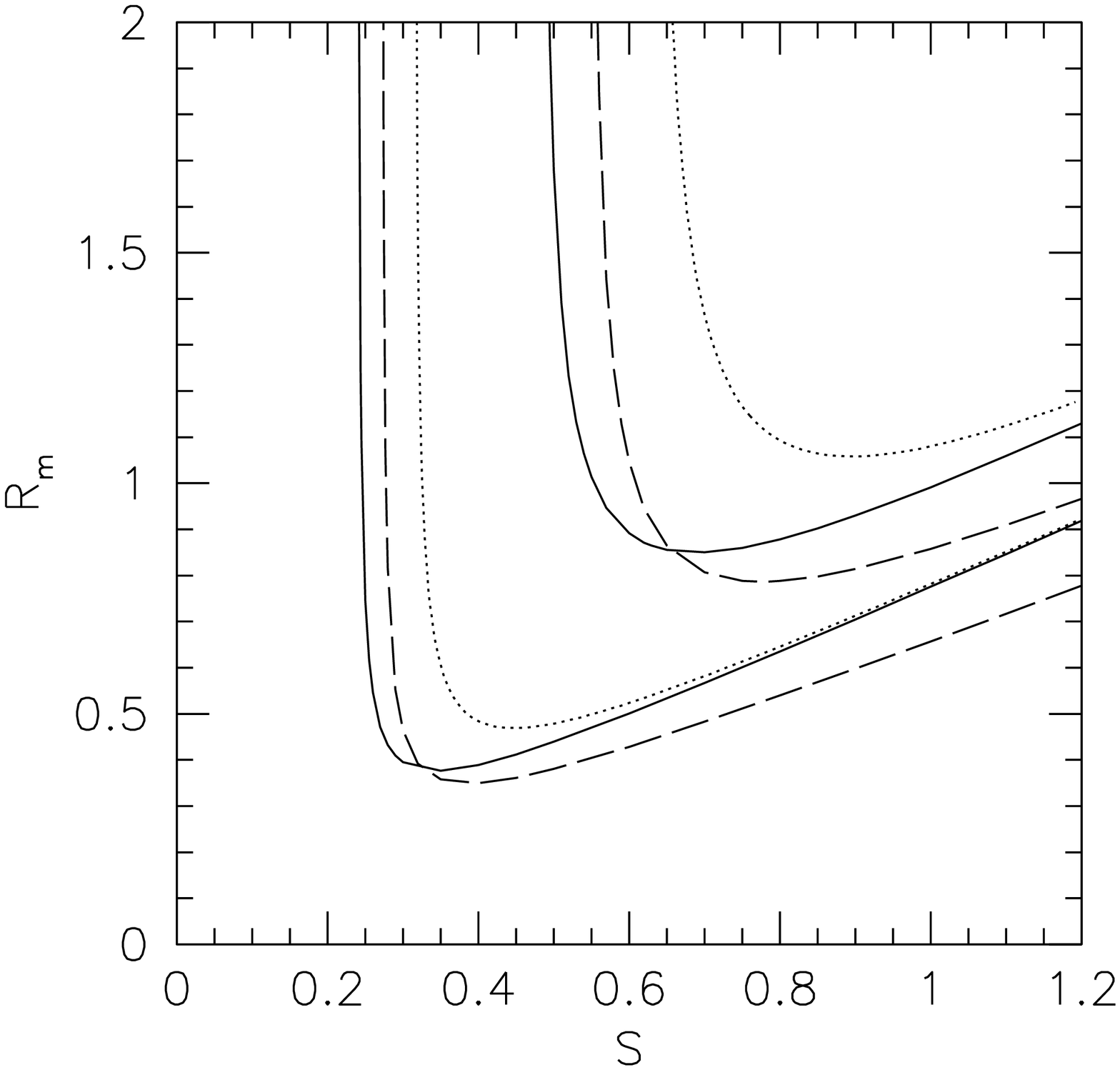}}
\end{minipage}
\hfill
\begin{minipage}{3in}
\scalebox{0.45}{\includegraphics{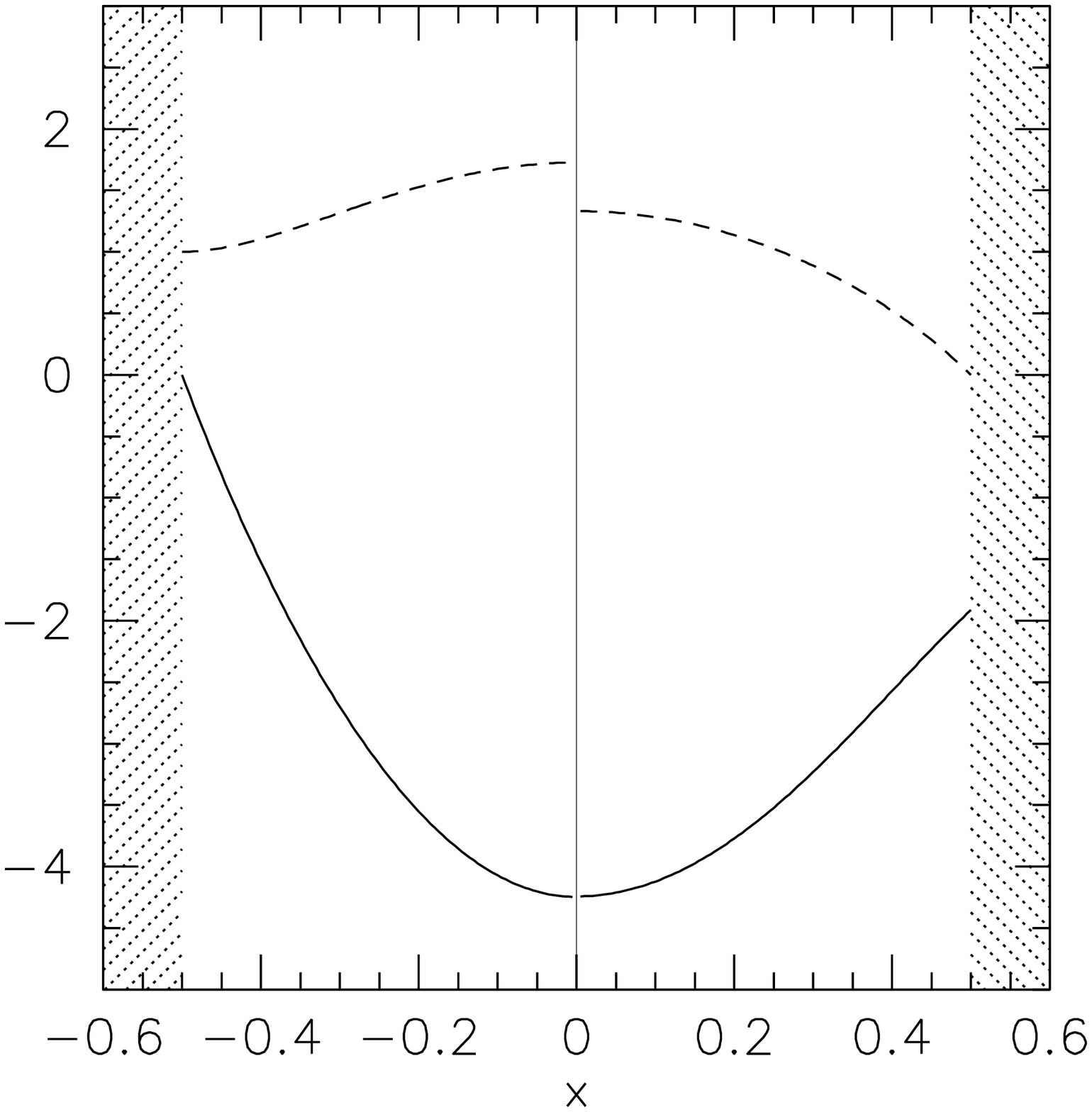}}
\end{minipage}
\caption{Narrow-gap modes for $\Pm=0$ and $\epsilon=1$.  {\it Left panel:}
Curves of marginal stability for $\zeta=2/11$ (lower curves)
and $\zeta=4/7$ (upper).  Solid line for conducting walls, dashed for
insulating, and dotted for local approximation (\ref{local}).
{\it Right panel:}
Narrow-gap eigenfunctions 
$\beta_r$ (solid curves) and $\beta_\theta$ (dashed), for $\zeta=2/11$
and $S=0.4$.
Since eigenfunctions are symmetric about center
of gap ($x=0$), only half of each is shown: conducting on left, 
insulating on right.  
\label{fig_narmode}}
\end{figure}

The narrow-gap limit $ A \to\infty$ is experimentally impractical but
theoretically important.
Fig.~\ref{fig_narmode} shows eigenfunctions and curves of marginal
stability and in this limit.  Because of the boundary conditions,
the eigenfunctions cannot be sinusoidal in $r$ or
$x\equiv (r-r_1)/d$, even though the equations of motion have
elementary solutions of this form.  The  fourth-order inviscid system
(\ref{inviscid1})-(\ref{inviscid2}) has four roots for the radial
wavenumber (or two if sign is ignored) at given parameters
$(\epsilon,S,\Rm,\zeta)$ of the equilibrium, which are constant across the gap.
(Note $K$ is not an independent parameter, since
$Kd=\sqrt{\epsilon^2+1}$.)
The solutions satisfying the boundary conditions are linear
combinations of four complex exponentials, each containing one of
these wavenumbers, and the parameters $(\epsilon,S,\Rm,\zeta)$ must
satisfy one constraint in order that a solution exist.
If, however, one simply sets $k_r=\pi/d$ in hopes of obtaining
an approximate constraint, then $D_k\to-K^2$ and
eqs.~(\ref{inviscid1})-(\ref{inviscid2}) or (\ref{fourthorder}) would
yield
\begin{equation}\label{local}
\Rm^2= S^2\frac{1+\epsilon^2}{2(2-\zeta-\zeta S^{-2})}.
\end{equation}
This corresponds to the dispersion relation for marginal modes
obtained in Paper I from a local WKB analysis with a ratio $\epsilon$ of
vertical to horizontal wavelength.  Evidently there exists a minimum
Lundquist number for instability,
\begin{equation}\label{Smin}
S_{\rm min} = \sqrt{\frac{\zeta}{2-\zeta}},
\end{equation}
and in the opposite limit of large $S$,
\begin{equation}\label{slope}
\frac{\Rm}{S}=\frac{\Omega}{kV_A}\approx \sqrt{\frac{1+\epsilon^2}{2(2-\zeta)}}.
\end{equation}
Fig.~(\ref{fig_narmode}) shows that the predictions
(\ref{local})-(\ref{slope}) are qualitatively correct.

Numerical growth rates are given in Table~1 for two representative
angular-velocity profiles, both with $r_2=3r_1=15\,\cm$, $h=10\,\cm$,
and the material properties of gallium.  
The first case has $\bar\zeta=0.063$ and hence is Rayleigh-stable.
The second has $\bar\zeta=-0.019$, so that the Rayleigh instability
occurs at zero field strength.  Growth rates are shown for conducting
and insulating radial boundary conditions.  
Larger fields are required to initiate and to quench MRI with
insulating boundaries than with conducting ones, presumably because
perturbed lines of force expand past insulating walls
into a volume slightly larger than the Couette flow itself.
The final column of this table shows growth rates computed from
the algebraic local dispersion relation given by in Paper I.  Once again
the WKB analysis predicts the global growth rates remarkably well,
even though the wavelengths involved are actually
larger than the gap width, and the angular velocity and shear
rate vary by a factor $\approx 9$ across the gap.

\begin{figure}
\hspace*{-0.4in}
\begin{minipage}{3.5in}
\scalebox{0.65}{\includegraphics{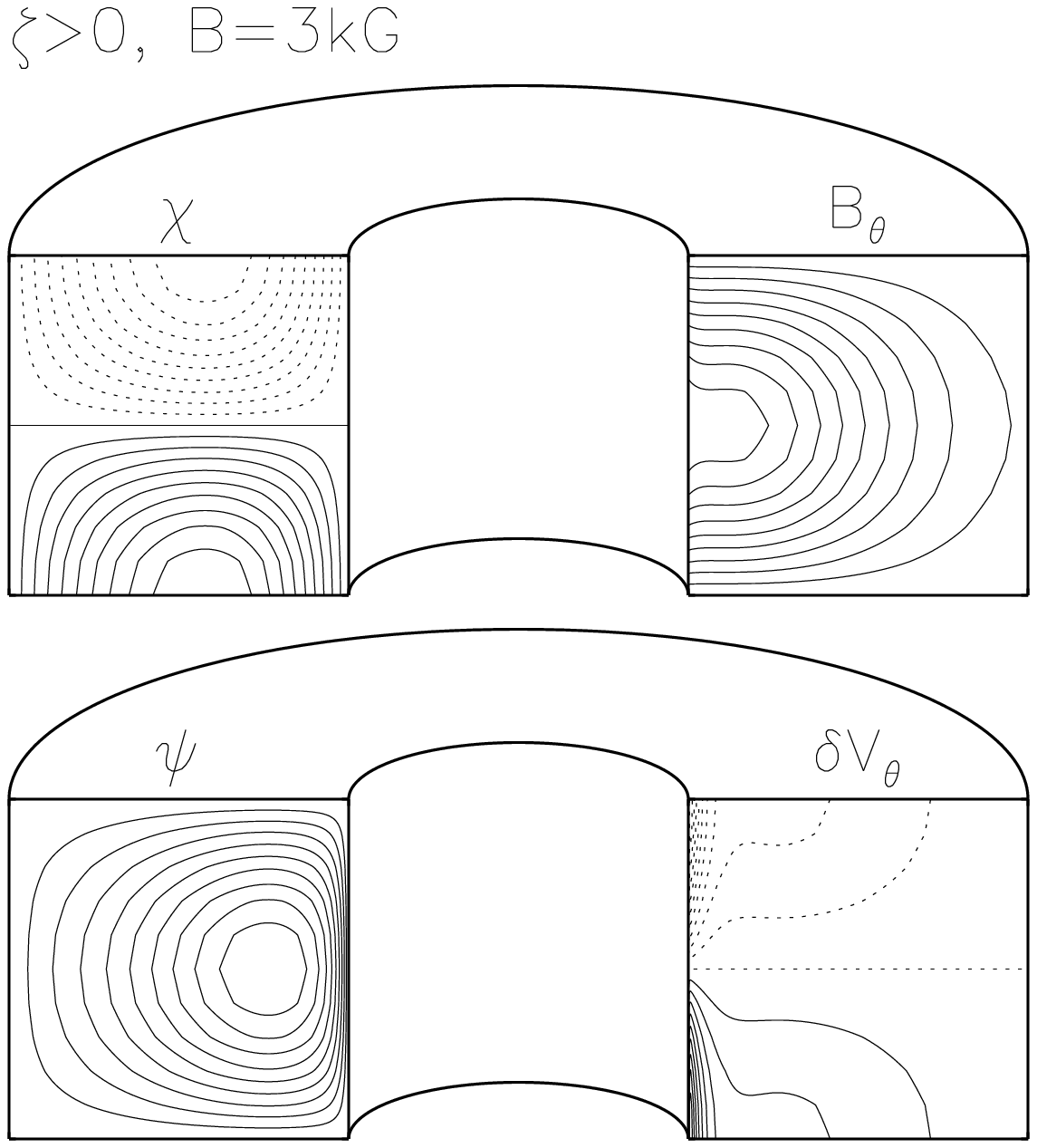}}
\end{minipage}
\begin{minipage}{3.5in}
\scalebox{0.65}{\includegraphics{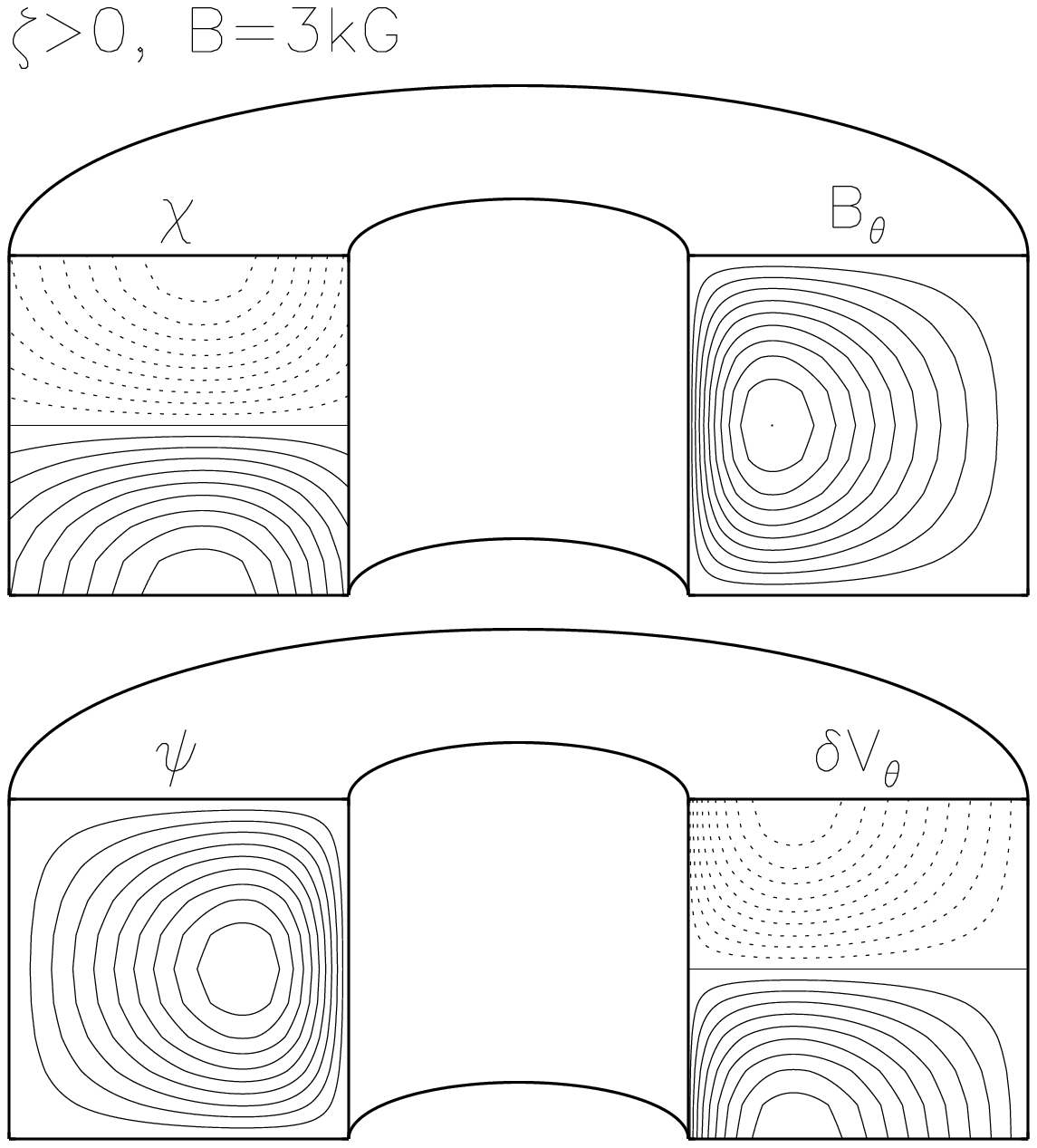}}
\end{minipage}
\caption{Visualizations of the MRI eigenmodes for the
Rayleigh-stable cases from Table 1 at $B_{z,0}=3$~kG.
{\it Left}: conducting boundaries.  {\it Right}: insulating.
Solid and dotted lines indicate positive and negative values, respectively.
See eq.~(\ref{chipsidef}) for definitions of
flux and stream functions $\chi,\psi$.
\label{fig_rs2D}}
\end{figure}

\begin{figure}
\hspace*{-0.4in}
\begin{minipage}{3.5in}
\scalebox{0.65}{\includegraphics{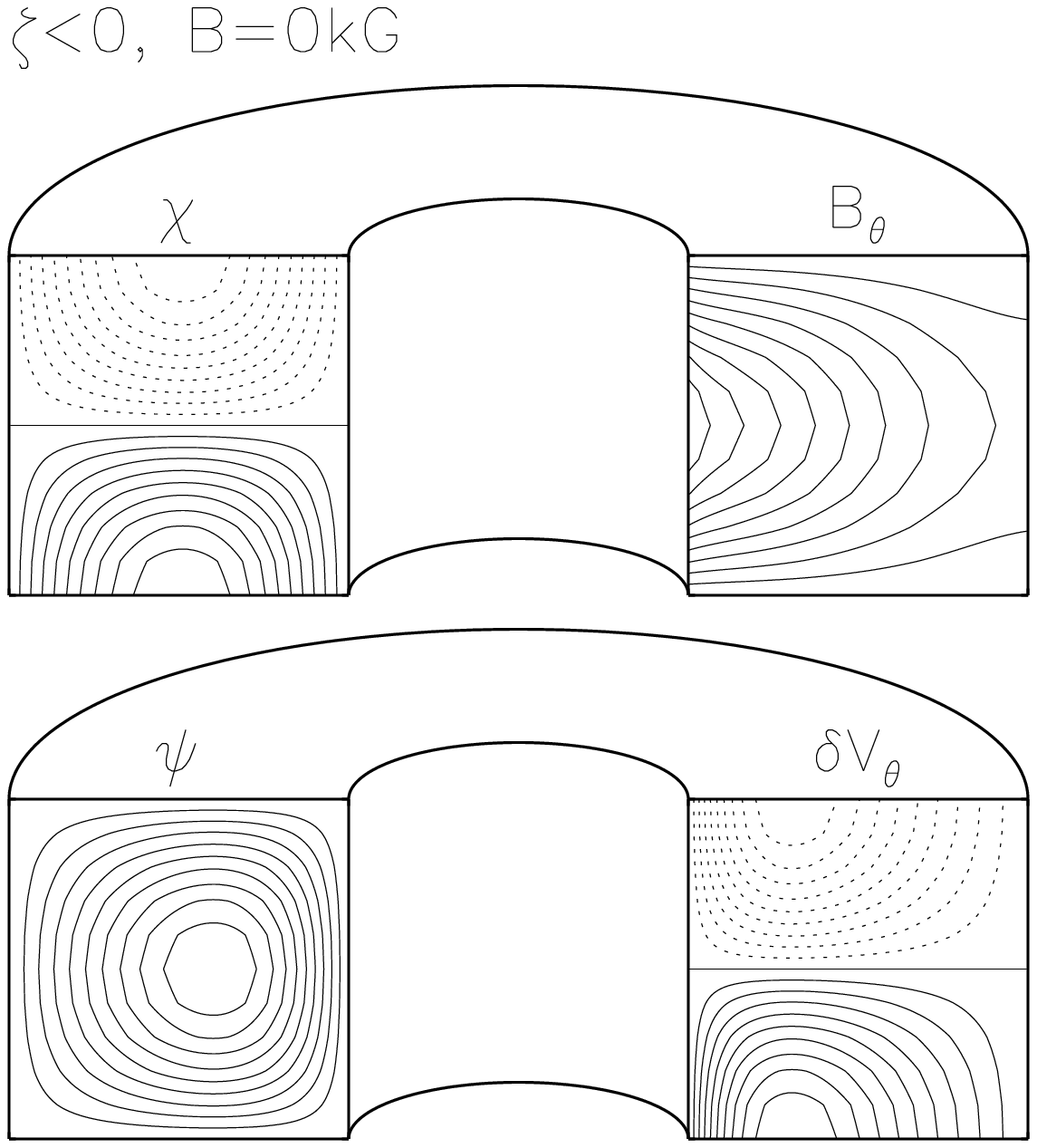}}
\end{minipage}
\begin{minipage}{3.5in}
\scalebox{0.65}{\includegraphics{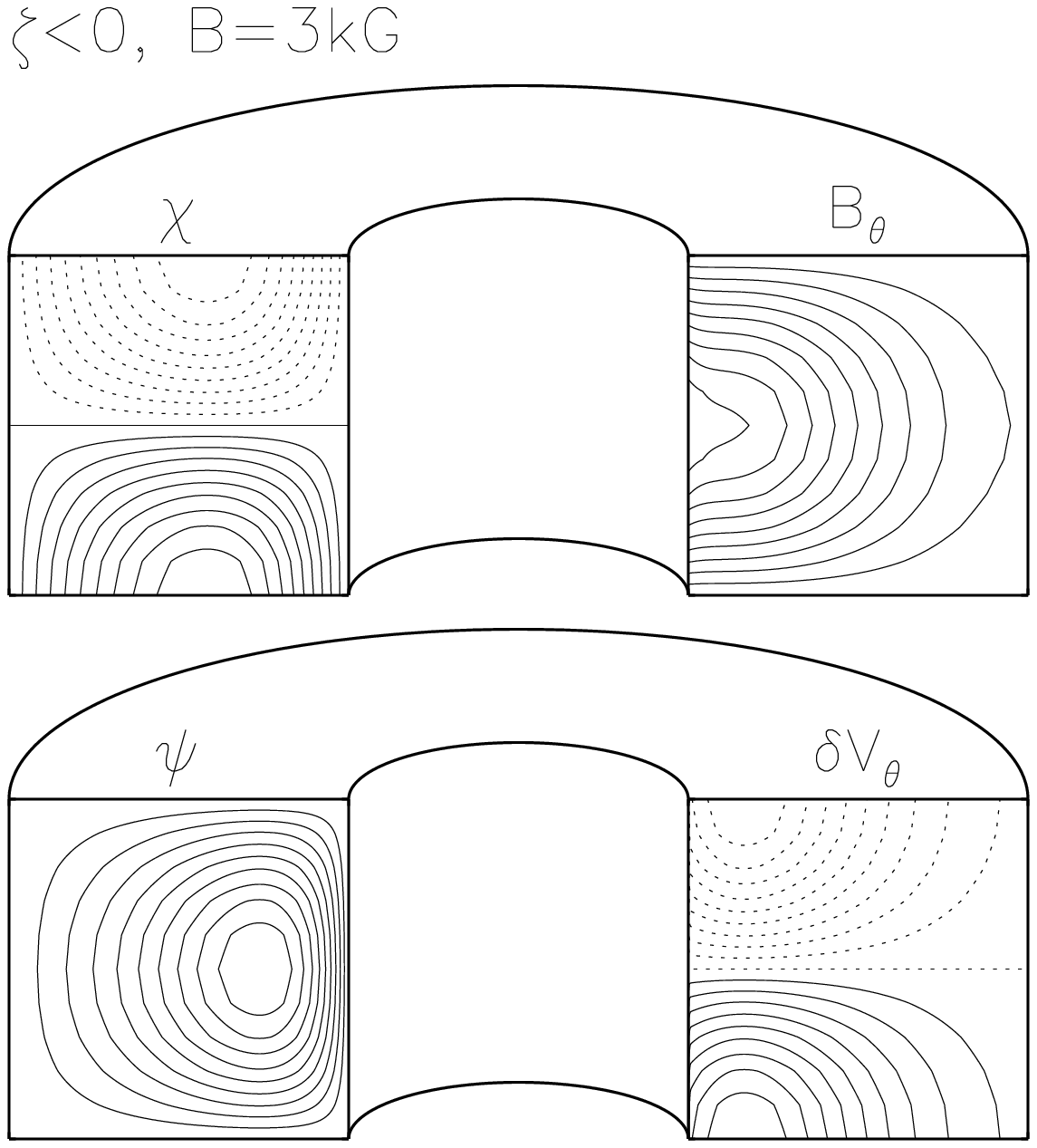}}
\end{minipage}
\caption{Rayleigh-unstable cases from Table 1 at
of $B_{z,0}=0$~kG (left) and 3~kG (right), both with 
conducting boundaries.
\label{fig_rus2D}}
\end{figure}

Figures (\ref{fig_rs2D}) and (\ref{fig_rus2D}) show two-dimensional
cross sections of selected modes from Table 1.  The flux and
stream functions are related to the poloidal perturbations by
\begin{equation}\label{chipsidef}
\frac{\delta\vB_P}{\sqrt{4\pi\rho}}=\grad\theta\times\grad\chi\,,\qquad
\delta\vv_P=\grad\theta\times\grad\psi\,.
\end{equation}
In all cases, the poloidal velocity field consists of a single roll.
The effect of the choice of boundary conditions is seen most
clearly in the toroidal perturbations.  In the first part of
Fig.~(\ref{fig_rs2D}) and the second part of Fig.~(\ref{fig_rus2D}),
it looks as though $\delta v_\theta$ does not vanish at the inner
boundary; in fact it does vanish, but the viscous boundary layer is
too small to be resolved by these plots.  
The first part of Fig.~(\ref{fig_rus2D}) does not show this behavior
because the magnetic forces are absent; this mode
is a classical hydromagnetic centrifugal instability.  A boundary layer
of the ordinary nonmagnetic variety occurs in $\delta v_z$.
The corresponding magnetized case shown in the second part of the figure
has a magnetically-driven boundary layer similar to that of
the Rayleigh-stable flows.

\section{Summary and discussion}

We have presented a global linear stability analysis for magnetized
Couette flow, including the dissipative effects of viscosity and
resistivity, in regimes where magnetorotational instability (MRI) is
possible.  In view of the actual properties of liquid metals, and
for plausible experimental lengthscales, 
resistivity is the main obstacle that must be overcome
to demonstrate MRI.  Previous theoretical studies
of magnetized Couette flow have focused on the 
problem of suppressing Rayleigh instability with a magnetic field,
and they have simplified the induction equations so
as to reduce the number of radial and time derivatives in the problem.
Such approximations may be adequate for the regime of interest to those
studies, but they will not do if one is interested in MRI.  Therefore
we have worked with the full induction equations.

Particular attention has been given to marginal stability.  All of our
numerical evidence indicates that exchange of stability occurs at
vanishing complex growth rates; we have not encountered any overstable
modes.  In the inviscid limit of marginal stability, the number of
radial derivatives reduces from ten to four, but the velocity
perturbations become singular at conducting boundaries.  The dominant
singularity arises from an unbalanced tangential magnetic force, not
from the usual pressure and inertial effects that cause boundary
layers in unmagnetized fluids.  Despite these complications, the
inviscid approximation predicts the locus of marginal stability
reasonably well for liquid metals.

Solving the linearized initial-value problem by finite-differences,
we have calculated growth rates and stability boundaries for a
liquid metal approximating gallium in an experimentally plausible
geometry.  For easier comparison with other theoretical work,
we have also made calculations in the narrow-gap limit
and expressed our results in the dimensionless coordinates of
Lundquist number and magnetic Reynolds number.

Remarkably, the growth rates are reasonably well
predicted by a simple WKB approximation even though the WKB modes
do not satisfy the boundary conditions and have a radial wavelength
twice the gap width, and even when the rotation rate varies by an
order of magnitude across the gap.  We conclude from this that the
algebraic WKB dispersion relation can be used for preliminary experimental
design, at least for aspect ratios no more extreme than considered
here ($r_2:r_1=3:1$).

There are good reasons to attempt an MRI experiment.  First,
one can hardly exaggerate the importance of this instability: few or
no plausible alternative explanations exist for the dissipation of
orbital energy in accretion disks, which are fundamental to so many of
the most energetic sources known in the universe.  Yet all present
knowledge of the instability is purely theoretical, based as it is on
linear analysis and computer simulation; the constraints provided by
astronomical observations are very indirect.  It is prudent to
put these theories to a laboratory test.

Secondly, computer speed limits the range of spatial scales that can
be modeled in the simulations.  Barring unforseen algorithmic
breakthroughs, the smallest resolvable scale in a three-dimensional
simulation improves only as the fourth root of the rate of arithmetic
operations.  Here it must be acknowledged that the large magnetic
diffusivity of liquid metals severely limits the number of degrees of
freedom in the magnetic field that can be excited.  The simulations
are well ahead of any forseeable experiment in this respect.  In fact,
simulations indicate that magnetic Reynolds numbers and Lundquist
numbers at least $100$ times larger than the minimum necessary for
linear instability are required for dynamo action in the absence of an
externally imposed field parallel to the rotation axis
\citep{FSH00}.  On the other hand, the viscous Reynolds number 
of such an experiment would be $R_{\rm e}\gtrsim 10^6$, a value still
out of reach of direct numerical simulations.
Also, small $\Rm$ need not restrict the experiment to linear behaviors.
In the local disk simulations of \cite{FSH00} at about twice the
minimum $\Rm$ for linear MRI, a violently fluctuating
nonlinear state was reached in which the time-averaged magnetic energy
was about 25 times larger than that of the externally imposed field.

Although large $\Rm$ is the rule in astrophysics,
the dimensionless parameters of some systems may be similar to
those of our proposed experiment, \emph{viz.} $\Rm$
and $S$ of order unity, $R_{\rm e}$ very large, and an externally imposed
field.  Such systems include the inner parts of relatively cool disks
(protostellar disks and quiescent cataclysmic variables, for example)
around stars with their own magnetic moments \citep{Gammie96, GM98}.

Lastly, relatively little laboratory MHD work has been done in
which the inertia of the fluid is important (large plasma $\beta$).
The experimental field appears somewhat underdeveloped when
measured against its potential importance to geophysics and astrophysics.
Because it promises to be achievable at fairly modest cost in a
classic experimental framework (Couette flow), MRI is a good
place to start.

\acknowledgements
This work was supported by the U.S. Department of Energy [H.J.]
and by NASA grant NAG5-8385 [J.G.]

\appendix
\section{Derivation of linearized equations}
For completeness, eqs.~(\ref{vsystem1})-(\ref{vsystem5}) are derived
here, although much the same derivation can be found in \cite{Chandra61}.
The equations of incompressible MHD are
\begin{eqnarray*}
\dot\vB+\vv\cdot\grad\vB-\vB\cdot\grad\vv&=&\eta\nabla^2\vB,\qquad\diver\vB=0,
\\
\dot\vv+\vv\cdot\grad\vv +\rho^{-1}\grad P
-\frac{\vB\cdot\grad\vB}{4\pi\rho}&=&\nu\nabla^2\vv,\qquad\diver\vv=0,
\end{eqnarray*}
in which $P\equiv p\,+\vB^2/8\pi$, is the hydrodynamic plus magnetic
pressure.
In cylindrical coordinates, near an equilibrium
$\vB_0=B\ve_z=\mbox{constant}$ and $\vv_0=r\Omega(r)\ve_\theta$,
linearized axisymmetric perturbations $\delta\vv$ and $\delta\vB$ satisfy
\begin{eqnarray}
\label{dbreqn}
\delta\dot B_r-B\pd_z\delta v_r&=&\eta(\pd_r\pd_r^\dag+\pd_z^2)\delta B_r,\\
\label{dbteqn}
\delta\dot B_\theta-B\pd_z\delta v_\theta -\delta B_r r\pd_r\Omega &=&
\eta(\pd_r\pd_r^\dag+\pd_z^2)\delta B_\theta,\\
\label{dvreqn}
\delta\dot v_r-2\Omega\delta v_\theta +\pd_r\frac{\delta P}{\rho}
-\frac{B}{4\pi\rho}\pd_z\delta B_r&=&\nu(\pd_r\pd_r^\dag+\pd_z^2)\delta v_r,\\
\label{dvteqn}
\delta\dot v_\theta+\delta v_r\pd_r^\dag(r\Omega)
-\frac{B}{4\pi\rho}\pd_z\delta B_\theta&=&
\nu(\pd_r\pd_r^\dag+\pd_z^2)\delta v_\theta,\\
\label{dvzeqn}
\delta\dot v_z+\pd_z\frac{\delta P}{\rho}
-\frac{B}{4\pi\rho}\pd_z\delta B_z&=&\nu(\pd_r^\dag\pd_r+\pd_z^2)\delta v_z,\\
\label{diverdB}
\pd_r^\dag\delta B_r +\pd_z\delta B_z &=&0,\\
\label{diverdv}
\pd_r^\dag\delta v_r +\pd_z\delta v_z &=&0,
\end{eqnarray}
in which the dot denotes $\pd/\pd t$, and other recurring operators are
\begin{displaymath}
\pd_z\equiv\frac{\pd}{\pd z}\,,\quad
\pd_r\equiv\frac{\pd}{\pd r}\,,\quad
\pd_r^\dag\equiv\frac{\pd}{\pd r}\,+\frac{1}{r}\,.
\end{displaymath}
Eqs.~(\ref{dvreqn}) \& (\ref{dvzeqn})
presume that $\rho$, like $\eta$ and $\nu$,
is spatially constant.
Applying $\pd_r^\dag$ to eq.~(\ref{dvreqn}) and
$\pd_z$ to eq.~(\ref{dvzeqn}) and summing the results, 
one finds that
\begin{displaymath}
(\pd_r^\dag\pd_r+\pd_z^2)\frac{\delta P}{\rho}=
\pd_r^\dag(2\Omega\delta v_\theta),
\end{displaymath}
in view of eqs.~(\ref{diverdB}) \& (\ref{diverdv}).
With another application of $\pd_r$, this becomes
\begin{equation}\label{Pieqn}
(\pd_r\pd_r^\dag+\pd_z^2)\Pi=\pd_z^2(2\Omega\delta v_\theta),\qquad
\mbox{where}~~\Pi\equiv 2\Omega\delta v_\theta - \pd_r\frac{\delta P}{\rho}~,
\end{equation}
so that the radial euler equation (\ref{dvreqn}) can be stated as
\begin{equation}
\label{dvreqn1}
\delta\dot v_r -\Pi -\frac{B}{4\pi\rho}\pd_z\delta B_r=
\nu(\pd_r\pd_r^\dag+\pd_z^2)\delta v_r.
\end{equation}
With the $z$ dependences given by eq.~(\ref{zdepend}) for the
linearized quantities, eqs.~(\ref{dbteqn}), (\ref{dvteqn}),
(\ref{dbreqn}), (\ref{dvreqn1}), and (\ref{Pieqn}) reduce to
eqs.~(\ref{vsystem1}), (\ref{vsystem2}), (\ref{vsystem3}),
(\ref{vsystem4}), and (\ref{vsystem5}), respectively.

{}



\newpage


 \begin{table}[t]
 \begin{tabular}{llll}
 \multicolumn{3}{c}{\bf Table 1.  Growth rates in gallium}\\ 
\hline\hline\\[10pt]
\multicolumn{4}{c}{\underline{$\Omega_1=413.6,~\Omega_2=~50.
~{\rm rad~s}^{-1}$}}
\\[10pt]
$B_z$ & \it conducting & \it insulating & \it local\\
 {[G]} & [s$^{-1}$] & [s$^{-1}$] & [s$^{-1}$] \\[5pt]
1893.&~0.00&	~--- & ~--- \\
2135.&	~5.35&	~0.00& ~--- \\
2500.&	~9.46&	~6.46& ~6.55 \\
3000.&	11.50&	10.83& 10.48 \\
3500.&	10.96&	12.66& 11.36\\
4000.&	~8.42&	12.84& ~9.99\\
4500.&	~4.16&	11.78& ~6.63\\
4868.&  ~0.00&	10.38& ~2.91\\
5500.&	~--- &	~7.10& ~---\\
6000.&	~--- &	~3.97& ~---\\
6588.&	~--- &	~0.00& ~---\\[20pt]
\multicolumn{4}{c}{\underline{$\Omega_1=377.0,~\Omega_2=
~40.84~[{\rm rad~s}^{-1}]$}}
\\[10pt]
$B_z$ & \it conducting & \it insulating & \it local\\
0. & 16.07 & 16.07 &   17.11\\
500. & 17.46 & 17.00&  17.98\\
1000. & 20.10 & 19.04& 19.85 \\
1500. & 22.23 & 21.07& 21.58 \\
2000. & 23.20 & 22.50& 22.49 \\
3000. & 21.13 & 22.83& 20.78 \\
4000. & 14.04 & 19.82& 13.50 \\
5000. & 4.219 & 14.38& ~---  \\
5500. & 1.438 & 11.22& ~---  \\
6000. & --- & 8.093  & ~--- \\
6500. & --- & 5.261  & ~--- \\
7000. & --- & 2.987  & ~--- \\
8000. & --- & .8246  & ~--- \\
9000. & --- & .4272  & ~--- \\
10000. & --- & .2131 & ~--- \\
11000. & --- & 3.614e-2 & ~--- \\
11220. & --- & 0.       & ~--- 
\end{tabular}
\end{table}

\end{document}